\documentclass[prl,twocolumn,superscriptaddress,showpacs]{revtex4-1}

\usepackage{amsmath}
\usepackage{amssymb}
\usepackage{graphicx}
\usepackage{color}
\usepackage{soul}

\begin{document}

\title{Gaussian benchmark for optical communication aiming towards ultimate capacity}

\author{Jaehak Lee}
\email{jaehak.lee@qatar.tamu.edu}
\affiliation{Department of Physics, Texas A \& M University at Qatar, P.O. Box 23874, Doha, Qatar}
\author{Se-Wan Ji}
\affiliation{Department of Physics, Texas A \& M University at Qatar, P.O. Box 23874, Doha, Qatar}
\author{Jiyong Park}
\affiliation{Department of Physics, Texas A \& M University at Qatar, P.O. Box 23874, Doha, Qatar}
\author{Hyunchul Nha}
\email{hyunchul.nha@qatar.tamu.edu}
\affiliation{Department of Physics, Texas A \& M University at Qatar, P.O. Box 23874, Doha, Qatar}
\affiliation{School of Computational Sciences, Korea Institute for Advanced Study, Seoul 130-722, Korea}

\begin{abstract}
We establish the fundamental limit of communication capacity within Gaussian schemes under phase-insensitive Gaussian channels, which employ {\it multimode} Gaussian states for encoding and {\it collective} Gaussian operations and measurements for decoding. We prove that this Gaussian capacity is additive, i.e., its upper bound occurs with separable encoding and separable receivers so that a single-mode communication suffices to achieve the largest capacity under Gaussian schemes. This rigorously characterizes the gap between the ultimate Holevo capacity and the capacity within Gaussian communication, showing that Gaussian regime is not sufficient to achieve the Holevo bound particularly in the low-photon regime. Furthermore the Gaussian benchmark established here can be used to critically assess the performance of non-Gaussian protocols for optical communication. We move on to identify non-Gaussian schemes to beat the Gaussian capacity and show that a non-Gaussian receiver recently implemented by Becerra {\it et al}. [Nat. Photon. {\bf 7}, 147 (2013)] can achieve this aim with an appropriately chosen encoding strategy.
\end{abstract}


\maketitle

{\it Introduction}---Sending and receiving signals via optical channels, e.g. optical fiber networks, is a crucial basis of communication. Employing protocols like intensity modulation and phase-shifting in optical communication \cite{bib:RevModPhys.58.1001,bib:PhysRevLett.70.363,bib:RevModPhys.66.481,bib:OpticalCommunication}, there eventually arises a question of fundamental importance---how quantum mechanics sets bound on communication capacity achievable using light beams. A remarkable result was recently established by proving the minimum output entropy conjecture \cite{bib:NatPhoton.8.796,bib:NatCommun.5.3826}, i.e., the ultimate capacity under phase-insensitive Gaussian channels is achieved by using coherent states as information carriers (encoding). However, there still exists an outstanding problem on what quantum receivers (decoding) can practically be used to obtain (near) ultimate capacity. 
The Holevo-Schumacher-Westmoreland theorem states that the ultimate capacity \cite{bib:ProblInfTransm.9.177} can be achieved {\it asymptotically} (using infinitely many channels) with a certain joint measurement \cite{bib:TransInf.44.269,bib:PhysRevA.56.131}, which however requires highly nonlinear, so very demanding, operations. It is therefore important to identify quantum receivers achieving high communication rates practically.

Numerous studies on quantum receivers focused mostly on distinguishing a finite set of coherent states with error rate below standard quantum limit (SQL) \cite{bib:Kennedy,bib:Dolinar,bib:Nature.446.774,bib:PhysRevA.78.022320,bib:NatPhoton.6.374}. For binary coherent states, the Dolinar receiver \cite{bib:Dolinar,bib:Nature.446.774} among them achieves the minimum error rate (Helstrom bound) \cite{bib:Helstrom}. Extending to $M$-ary signals ($ M > 2 $), several receivers have also been proposed \cite{bib:OptLett.18.1896,bib:NewJPhys.14.083009,bib:PhysRevA.84.062324,bib:PhysRevA.86.042328,bib:PhysRevA.89.032318} and experimentally demonstrated e.g. by adaptive phase nulling and photon counting \cite{bib:NatPhoton.7.147,bib:NatPhoton.9.48}. 
While the performance of these receivers was evaluated by error rate below SQL under specific codewords \cite{bib:NatPhoton.6.374,bib:NatPhoton.9.48},  
it is critically important to see how those {\it non-Gaussian} receivers manifest advantages in terms of capacity (mutual information). 
We thus need to identify the capacity achievable within {\it Gaussian communication schemes} employing Gaussian states, operations, and measurements \cite{bib:PhysRevA.66.032316} readily available in laboratory \cite{bib:RevModPhys.84.621}. It is unknown to what extent general Gaussian schemes particularly using entangling operations can improve capacity in contrast to separable schemes.

So far there are two well-known Gaussian communication schemes, coherent-state scheme with heterodyne detection and squeezed-state scheme with homodyne detection, studied under an ideal situation \cite{bib:RevModPhys.58.1001,bib:RevModPhys.66.481} or channel noises \cite{bib:PhysRevA.50.3295,bib:OptCommun.149.152}. We recently extended study to general {\it single}-channel Gaussian communications with arbitrary inputs and measurements and showed that the optimal strategy among them is either coherent-state scheme or squeezed-state scheme depending on channel parameters \cite{bib:PhysRevA.91.042336}. As for multimode scenario, with inputs restricted to coherent states under thermal dissipative channels, Takeoka and Guha showed that the optimal Gaussian receiver is a separable one \cite{bib:PhysRevA.89.042309}. Since Gaussian receivers with coherent-state inputs do not saturate the ultimate channel capacity although the channel capacity is obtained with coherent-state inputs, their work provides an evidence for the gap between the capacity of Gaussian schemes and the ultimate channel capacity. However, the restriction to coherent-state inputs is not sufficient as other inputs (squeezed state) can yield higher capacity under some Gaussian channels \cite{bib:PhysRevA.91.042336}. 

In this paper we establish the ultimate limit of Gaussian schemes under phase-insensitive Gaussian channels in a general multimode scenario using arbitrary $N$-mode Gaussian input states and collective Gaussian measurements. We prove that its upper bound is achieved by separable inputs and separable measurements (additivity of Gaussian communication). The highest capacity of Gaussian schemes is thus obtained by the optimal {\it single-channel} protocol, i.e., either coherent-state scheme or squeezed-state scheme \cite{bib:PhysRevA.91.042336}. As the capacities of those two schemes do not achieve the Holevo bound \cite{bib:PhysRevA.50.3295,bib:PhysRevA.91.042336}, we characterize the exact gap between the ultimate channel capacity and the capacity within Gaussian communication.
Our results shed light on optical communication in several aspects. First, it identifies an optimal protocol when resources are confined to Gaussian operations and Gaussian receivers. 
Until now, coherent-state and squeezed-state schemes were used as standard protocols due to simple applicability. We now show that they actually attain the upper limit of capacity within Gaussian resources.
Second, it establishes a benchmark to rigorously assess enhanced performance of non-Gaussian schemes in terms of mutual information---a central quantity of interest in communication theory. Furthermore, we suggest a non-Gaussian receiver of \cite{bib:PhysRevA.84.062324,bib:NatPhoton.7.147} combined with an appropriate encoding method as a feasible scheme for higher communication rate than Gaussian limit.

{\it Gaussian Communication}---An $N$-mode Gaussian state is fully characterized by its first moments (averages) and second moments (variances) of position and momentum operators $ \hat{\boldsymbol{\xi}} = \left( \hat{x}_1,\hat{p}_1,\hat{x}_2,\hat{p}_2,\cdots,\hat{x}_N,\hat{p}_N \right)^T $ \cite{bib:RevModPhys.84.621}. The second-order moments are given by a covariance matrix (CM) $ \boldsymbol{\gamma} $ with elements $ \gamma_{ij} = \frac{1}{2} \left\langle \hat{\xi}_i \hat{\xi}_j + \hat{\xi}_j \hat{\xi}_i \right\rangle - \left\langle \hat{\xi}_i \right\rangle \left\langle \hat{\xi}_j \right\rangle $ ($i,j=1,\cdots,2N$). Let Alice prepare an $N$-mode Gaussian state $\rho_0$ with mean values  $\langle \hat{\xi}_i\rangle=0$ and CM $\boldsymbol{\gamma}_\textrm{in}$. She encodes random variables $ \boldsymbol{d}_\alpha = \sqrt{2}(\textrm{Re}\alpha_1,\textrm{Im}\alpha_1,\textrm{Re}\alpha_2,\textrm{Im}\alpha_2,\cdots,\textrm{Re}\alpha_N,\textrm{Im}\alpha_N)^T $ by performing displacements as $ \rho_\textrm{in} = \bigotimes_{j=1}^{N}\hat{D}_j(\alpha_j) \rho_0 \bigotimes_{j=1}^{N}\hat{D}_j^\dagger(\alpha_j) $, with $ \hat{D}_j(\alpha) = \exp\left( \alpha\hat{a}_j^\dagger - \alpha^*\hat{a}_j \right) $  a displacement operator on $j$th mode. While Gaussian schemes use Gaussian operations and measurements, no restriction is given to the probability distribution $P(\boldsymbol{d}_\alpha)$ of encoded variables (continuous or discrete). In the classical information theory \cite{bib:InformationTheory}, it is known that if the added noise is Gaussian, the input distribution must also be Gaussian to optimize capacity. In our Gaussian communication scenario, both of the noise emerging from the internal fluctuation of output states and the noise induced by measurement are Gaussian. Therefore, mutual information is maximized by a Gaussian distribution 
	 \begin{equation} \label{eq:probencoding}
	P(\boldsymbol{d}_\alpha) = \frac{1}{(2\pi)^N\sqrt{\det P_\textrm{in}}} \exp\left( -\frac{1}{2} \boldsymbol{d}_\alpha^{T} P_\textrm{in}^{-1} \boldsymbol{d}_\alpha \right) ,
	\end{equation}
with $P_\textrm{in}$ a $2N\times2N$ real positive matrix characterizing the range (variance) of encoded variables.

Alice sends the $N$-mode state to Bob via $N$ independent Gaussian channels, which gives an output $\rho_\textrm{out}$ with amplitudes $ \boldsymbol{d}_\alpha' = \sqrt{\tau}\boldsymbol{d}_\alpha $ 
and CM
	\begin{equation} \label{eq:cmchannel}
	\boldsymbol{\gamma}_\textrm{out} = T \boldsymbol{\gamma}_\textrm{in} T^{T} + M ,
	\end{equation}
where $ T = \sqrt{\tau}I^{\otimes N} $ and $ M = mI^{\otimes N} $ ($I$: $2\times2$ identity matrix). Note that an arbitrary phase-insensitive Gaussian channel can be understood as a concatenation of loss and amplification channels, fully characterized by two parameters $\tau$ and $m$ satisfying $m\ge|\tau-1|$\cite{bib:PhysRevLett.108.110505}. 
 For instance, a loss channel with transmittance $\eta\le1$ and thermal photons $n_\textrm{th}$ is characterized by $ \tau = \eta $ and $ m = (1-\eta)(n_\textrm{th}+\frac{1}{2}) $, while an amplification channel with gain $g\ge1$ is by $ \tau = g $ and $ m = (g-1)\left(n_\textrm{th}+\frac{1}{2}\right) $.
 
Bob finally obtains outcomes $ \boldsymbol{d}_\beta $ using an $N$-mode Gaussian measurement. An arbitrary $N$-mode Gaussian measurement can be described by a projection onto a Gaussian state $\rho_M$ with mean values $ \boldsymbol{d}_\beta $ and CM $\boldsymbol{\gamma}_M$.
The conditional probability for $\boldsymbol{d}_\beta$, given an input $\boldsymbol{d}_\alpha$, then follows as a Gaussian distribution centered at $\boldsymbol{d}_\alpha'$ with its second moments determined by both the internal fluctuation of $\rho_\textrm{out}$ and the added noise from $\rho_M$, 
	\begin{eqnarray}
	& & P(\boldsymbol{d}_\beta|\boldsymbol{d}_\alpha) = \frac{1}{(2\pi)^N\sqrt{\det (\boldsymbol{\gamma}_\textrm{out}+\boldsymbol{\gamma}_M)}} \\
	& & \qquad \times \exp\left[ -\frac{1}{2} (\boldsymbol{d}_\beta-\boldsymbol{d}_\alpha')^{T} (\boldsymbol{\gamma}_\textrm{out}+\boldsymbol{\gamma}_M)^{-1} (\boldsymbol{d}_\beta-\boldsymbol{d}_\alpha') \right] . \nonumber
	\end{eqnarray}
Additionally, one might construct a receiver employing partial measurement on some (ancillary) modes and classical feedforward. However, the feedforward scheme does not make improvement to our general Gaussian scenario, like the case of coherent-state input \cite{bib:PhysRevA.89.042309}, since any Gaussian measurement on Gaussian states can be transformed into a deterministic Gaussian operation \cite{bib:PhysRevA.66.032316,bib:PhysRevLett.89.137903,bib:PhysRevLett.89.137904}. It thus suffices to consider our settings without ancillary modes or feedforward scheme.

As for the calculation of capacity, our case resembles the classical Gaussian communication where signal power is given by $P_\textrm{out}= \tau P_\textrm{in} $ and noise by $\boldsymbol{\gamma}_\textrm{out}+\boldsymbol{\gamma}_M$ \cite{bib:InformationTheory}. The mutual information measured in bits between Alice and Bob is given by
	\begin{eqnarray} \label{eq:mi}
	I(A:B) & = & \int d^{2N}\boldsymbol{d}_\alpha d^{2N}\boldsymbol{d}_\beta ~ P(\boldsymbol{d}_\beta,\boldsymbol{d}_\alpha) \log_2 \frac{P(\boldsymbol{d}_\beta|\boldsymbol{d}_\alpha)}{P(\boldsymbol{d}_\beta)} \nonumber \\
	& = & \frac{1}{2} \log_2\frac{\det(P_\textrm{out}+\boldsymbol{\gamma}_\textrm{out}+\boldsymbol{\gamma}_M)}{\det(\boldsymbol{\gamma}_\textrm{out}+\boldsymbol{\gamma}_M)} .
	\end{eqnarray}
It grows indefinitely with the signal power $P_\textrm{out}$, so energy constraint is introduced for a realistic consideration. The average photon number per channel is bounded by $\bar{n}$, i.e. $ n_0 + n_s \le N\bar{n} $ where $ n_0 = \frac{1}{2}\left( \textrm{tr}\boldsymbol{\gamma}_\textrm{in}-N \right) $ is from input state and $ n_s = \frac{1}{2}\textrm{tr}P_\textrm{in} $ from signal encoding.
	
Without loss of generality, we only consider a pure-state input (encoding) and a projection onto a pure state (decoding). For a mixed state $\rho_\textrm{in}$ or $\rho_M$, it is always possible to find a pure state leading to a higher capacity. Any pure Gaussian state can be obtained by applying on a vacuum state a unitary operation decomposed as $USV$, with $U$ and $V$ passive transformations (energy conserving) and $S$ single-mode squeezing operations \cite{bib:PhysRevA.71.055801}. Using $ S_V S_V^T = I $ with $S_V$ a symplectic matrix for a passive unitary operation $V$ \cite{bib:PhysRevA.49.1567}, we can diagonalize CMs as
	\begin{equation} \label{eq:cmin}
	\boldsymbol{\gamma}_\textrm{in} = S_{U_0} \boldsymbol{\gamma}_\textrm{in}^{(D)} S_{U_0}^T
		\equiv S_{U_0} \bigoplus_{j=1}^{N} \left(\begin{array}{cc} \frac{1}{2}e^{-2r_j} & 0 \\ 0 & \frac{1}{2}e^{2r_j} \end{array}\right) S_{U_0}^T ,
	\end{equation}
	\begin{equation} \label{eq:cmmeasure}
	\boldsymbol{\gamma}_\textrm{M} = S_{U_M} \boldsymbol{\gamma}_{M}^{(D)} S_{U_M}^T
		\equiv S_{U_M} \bigoplus_{j=1}^{N} \left(\begin{array}{cc} \frac{1}{2}e^{-2s_j} & 0 \\ 0 & \frac{1}{2}e^{2s_j} \end{array}\right) S_{U_M}^T .
	\end{equation}
Here $S_{U_0}$ and $S_{U_M}$ are symplectic matrices for passive unitary operations $U_0$ and $U_M$, and $r_j$ and $s_j$ single-mode squeezing parameters. Using Eqs. (\ref{eq:cmchannel}) and (\ref{eq:cmin}), the output CM is diagonalized as
	\begin{equation}
	\boldsymbol{\gamma}_\textrm{out} = S_{U_0} \left( T\boldsymbol{\gamma}_\textrm{in}^{(D)}T^T + M \right) S_{U_0}^T = S_{U_0} \boldsymbol{\gamma}_\textrm{out}^{(D)} S_{U_0}^T ,
	\end{equation}
with the same $S_{U_0}$ as applying on the input. We first investigate the mutual information with fixed $r_j$ and $s_j$ in decreasing order $ r_1 \ge r_2 \ge \cdots \ge r_N \ge 0 $ and $ s_1 \ge s_2 \ge \cdots \ge s_N \ge 0 $ and may later optimize $r_j$ and $s_j$.

\begin{figure}
	\centering \includegraphics[width=0.5\columnwidth]{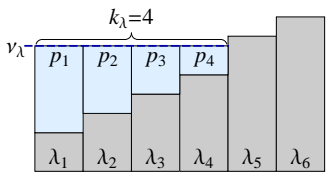}
	\caption{\label{fig:waterlambda} Illustration of water-filling solution (optimal encoding) for a given noise vector $\vec{\lambda}$.}
	\end{figure}

{\it Optimization}---For a given noise matrix $ \boldsymbol{\gamma}_\textrm{out}+\boldsymbol{\gamma}_M$, the mutual information (\ref{eq:mi}) can be maximized when signal $P_\textrm{out}$ is given by the so-called {\it water-filling} solution known in classical information theory \cite{bib:InformationTheory} (Fig. \ref{fig:waterlambda}).
Let $ \vec{\lambda} \equiv \left( \lambda_1, \lambda_2, \cdots, \lambda_{2N} \right)^T $ be the eigenvalues (not sympletic ones) of $ \boldsymbol{\gamma}_\textrm{out}+\boldsymbol{\gamma}_M $ in increasing order $ \lambda_1 \le \lambda_2 \le \cdots \le \lambda_{2N} $, with its diagonalization $ \boldsymbol{\Lambda} = \textrm{diag}\left( \lambda_1, \lambda_2, \cdots, \lambda_{2N} \right)=R^{-1}\left( \boldsymbol{\gamma}_\textrm{out}+\boldsymbol{\gamma}_M \right)R$ ($R$: similarity transformation) \cite{bib:PhysRevA.49.1567}. The mutual information is then maximized when $P_\textrm{out}$ is diagonalized by the same transformation, i.e., $ R^{-1}P_\textrm{out}R = \textrm{diag}\left( p_1, p_2, \cdots, p_{2N} \right) $. The optimal signal power $p_j$ assigned against the noise $\lambda_j$ is given by $ p_j = \textrm{max}\left\{ \nu_\lambda-\lambda_j, 0 \right\} $, where the ``water" level $\nu_\lambda$ is determined to saturate the energy constraint, $ \sum_{j=1}^{2N} \textrm{max}\left\{ \nu_\lambda-\lambda_j, 0 \right\} = \textrm{tr}P_\textrm{out} = 2\tau n_s = 2\tau ( N\bar{n} -n_0 ) $. Let $k_\lambda$ denote the number of nonzero signals i.e. $ \lambda_j < \nu_\lambda $ for $ j \le k_\lambda $ and $ \lambda_j \ge \nu_\lambda $ for $ j > k_\lambda $. Then, the mutual information via water-filling solution is given by
	\begin{eqnarray} \label{eq:milambda}
	&&I(A:B) = \frac{1}{2}\sum_{j=1}^{2N} \log_2\left( 1 + \frac{p_j}{\lambda_j} \right) = \frac{1}{2}\sum_{j=1}^{k_\lambda} \log_2\frac{\nu_\lambda}{\lambda_j} \nonumber \\
	& &= \frac{1}{2}\sum_{j=1}^{k_\lambda} \log_2\frac{2\tau n_s + \sum_{i=1}^{k_\lambda}\lambda_i}{k_\lambda\lambda_j} \equiv f\left( \lambda_1, \lambda_2 \cdots, \lambda_{k_\lambda} \right).
	\end{eqnarray}

In \cite{supple}, we prove that $ S_{U_0} = S_{U_M} = I $, i.e. CMs already diagonalized $\boldsymbol{\gamma}_\textrm{in} =\boldsymbol{\gamma}_\textrm{in}^{(D)}$ and $\boldsymbol{\gamma}_{M} =\boldsymbol{\gamma}_{M}^{(D)}$ representing product Gaussian states, gives the maximum in Eq. (\ref{eq:milambda}). Our proof relies on the majorization theory \cite{bib:Majorization,bib:MatrixAnalysis} and the fact that $ f\left( \lambda_1, \lambda_2 \cdots, \lambda_{k_\lambda} \right) $ in Eq. (\ref{eq:milambda}) is Schur-convex. It shows that a separable encoding/decoding is the optimal strategy, reducing the problem to finding the optimal {\it single-channel} Gaussian scheme.

\begin{figure}[t]
	\centering \includegraphics[width=0.9\columnwidth]{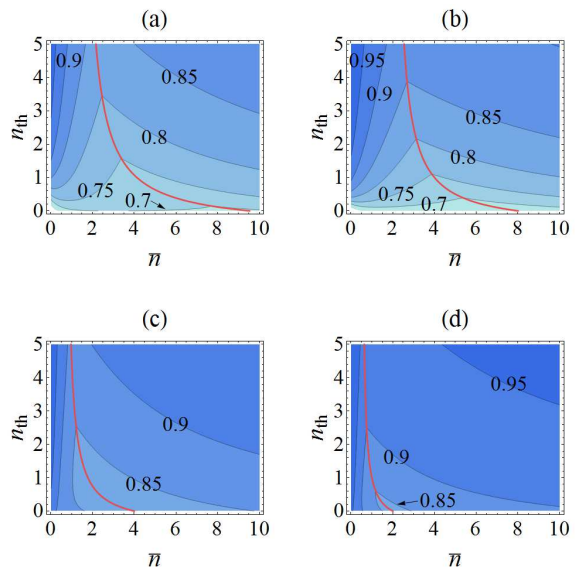}
	\caption{\label{fig:efficiency} Ratio of the Gaussian communication capacity to the ultimate Holevo bound against input energy $\bar n$ and thermal noise $n_{\rm th}$ under a lossy channel with (a) $\tau=0.7$, (b) $\tau=0.5$, and an amplification channel with (c) $\tau=1.5$ and (d) $\tau=2$. The red curves represent the crossover from $C^\textrm{sq}$ (left region) to $C^\textrm{coh}$ (right region) for the optimal Gaussian capacity.}
	\end{figure}

{\it Single-channel communication}---In \cite{bib:PhysRevA.91.042336}, we proved that the optimal Gaussian communication under a single-channel use is one of two well-known Gaussian schemes, coherent-state scheme or squeezed-state scheme depending on channel parameters \cite{note:optimalsingle}. In \cite{supple}, we give details for their capacities, $C^\textrm{coh}$ (coherent-state scheme) and $C^\textrm{sq}$ (squeezed-state scheme), compared with the ultimate Holevo bound $C^\textrm{Holevo}$ for completeness. Precisely, the Gaussian capacity approaches the Holevo bound in the high-energy limit, but does not so in low-photon regime of practical importance. 
Nevertheless, it attains a considerably high capacity in broad parameter regions. In Fig. \ref{fig:efficiency}, we plot the ratio of Gaussian communication capacity to the Holevo capacity, $ \textrm{max}\left\{C^\textrm{coh},C^\textrm{sq}\right\}/C^\textrm{Holevo} $.
With input energy $\bar n$ increasing, there generally occurs a crossover from squeezed-state scheme to coherent-state scheme for the highest Gaussian capacity, with critical $\bar{n}_c=\frac{1+2m+\tau}{2m\tau} $. For $n<n_c$, the squeezed-state scheme achieves a high efficiency ($>90\%$) for small $\bar{n}$ and large $n_\textrm{th}$. As $\bar{n}$ increases, the efficiency of squeezed-state scheme drops significantly and the optimal strategy turns into the coherent-state scheme for $n>n_c$. The coherent-state scheme becomes efficient with $\bar{n}$ and $n_\textrm{th}$ increasing. Under amplification channel [Fig. \ref{fig:efficiency}(c,d)], a high efficiency ($>90\%$) is achieved broadly with gain $g$ increasing. Under loss channel [Fig. \ref{fig:efficiency} (a,b)], such a high efficiency hardly appears with $n_\textrm{th}$ small.
Aiming at $80\%$ $(90\%)$ efficiency under pure-loss channel ($n_{\rm th}$=0), a large input energy $ \tau \bar{n} \gtrsim 52$ (8098) is required.

{\it Beyond Gaussian limit}---We have established the Gaussian benchmark for the capacity limit of Gaussian communication under general Gaussian settings, 
which turns out to be below the ultimate Holevo capacity particularly in the low-photon regime. It is then interesting to identify non-Gaussian protocols to beat this Gaussian limit.
A simple non-Gaussian receiver using photon counting was proposed for binary coherent inputs beyond Gaussian limit in the extremely low photon-number regime \cite{bib:PhysRevA.89.042309}. Because a binary input carries at most one bit of information, we extend to an $M$-ary signal modulation ($ M > 2 $) for higher communication rate. While some studies were done for $M$-ary signals \cite{bib:OptLett.18.1896,bib:NewJPhys.14.083009,bib:PhysRevA.84.062324,bib:PhysRevA.86.042328,bib:PhysRevA.89.032318,bib:NatPhoton.7.147,bib:NatPhoton.9.48}, the analysis was made on error rate against SQL. In contrast, capacity in bits is an important quantity in communication and we are now able to rigorously assess merits of non-Gaussian receivers against the Gaussian-capacity benchmark. 

We here investigate a non-Gaussian receiver recently demonstrated by Becerra {\it et al.} \cite{bib:PhysRevA.84.062324,bib:NatPhoton.7.147} using on-off detection without photon-number resolving for easy implementation, together with quadrature amplitude modulation (QAM) of coherent states as encoding. 
The Becerra receiver takes recursive steps of (i) splitting the coherent signal into $L$ pulses and (ii) carrying out signal nulling (via inverse displacement) and photon detection sequentially. 
For each pulse, the receiver chooses a most likely hypothetical coherent state among inputs to displace input to a vacuum and then detects photons. If the hypothesis is correct, no click occurs at the detector. Looking into the detection history of previous stages, the receiver updates the hypothesis via Bayesian conditional probability, determining the displacement at next stage. The whole detection outcomes are used to guess an input. 

\begin{figure}[!t]
	\centering \includegraphics[width=0.9\columnwidth]{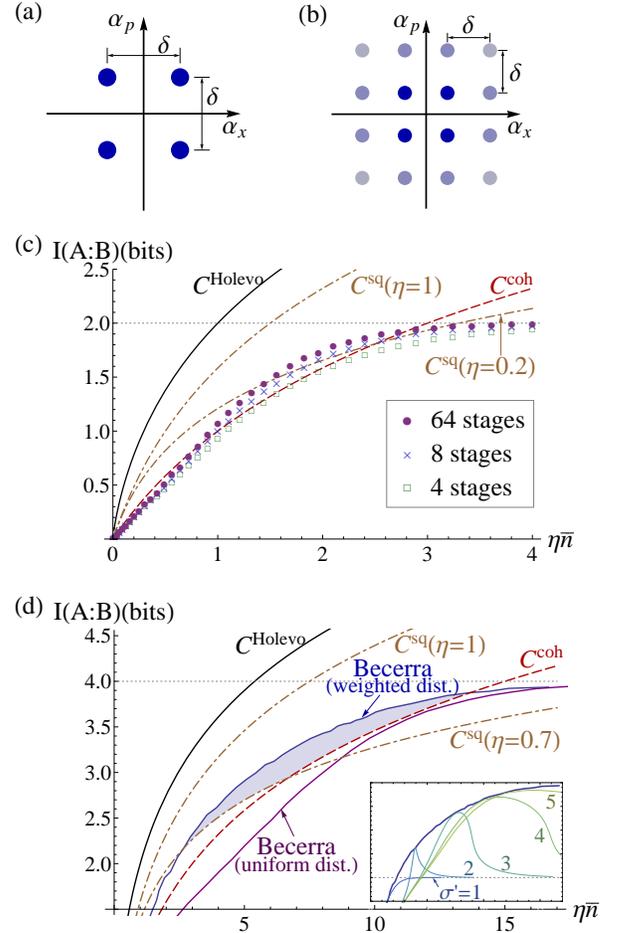}
	\caption{\label{fig:Becerra} (a) 4-QAM and (b) 16-QAM with more probable inputs in darker blue. Mutual information with (c) 4-QAM and (d) 16-QAM using the Becerra receiver, which beats all Gaussian schemes in the shaded region. Inset: capacity of Becerra receiver with $ \sigma' = \{1,2,3,4,5\} $ (thin curves) and optimal $\sigma$ (thick curve). Horizontal dotted line represents the maximum possible capacity 2-bits (4-bits) for 4-QAM (16-QAM).}
	\end{figure}

Under QAM encoding, Alice prepares one of the coherent states at lattice points with spacing $\delta$ in phase space (Fig. \ref{fig:Becerra}(a,b)).
In Fig. \ref{fig:Becerra}(c), we show the mutual information attained by the Becerra receiver for 4-QAM under a loss channel, compared with the Gaussian capacity limit and the Holevo bound \cite{note:QAMtransmission}.  Its capacity is improved by increasing the number of stages $L$. While it cannot beat the coherent-state scheme with $L=4$, it slightly does so with $L=16$ and 64. However, it is hard to beat the squeezed-state scheme, which is highly efficient for a large $\eta$ and a small $\bar{n}$. For a small transmittance $\eta \lesssim 0.2 $, the Becerra receiver begins to beat both Gaussian communication schemes.

To find enhanced performance of this non-Gaussian scheme by increasing the number of inputs, we move on to 16-QAM. With the pulse splitting to $L=64$, we find that the Becerra receiver cannot beat the Gaussian limit with a standard 16-QAM where all input states are prepared with uniform probability (Fig. \ref{fig:Becerra}(d)). This is because a coherent state with a larger displacement carries more energy, which needs to be probabilistically suppressed as we evaluate capacity under energy constraint. We thus investigate a modified 16-QAM where an input state is prepared according to a Gaussian-like distribution of deviation $\sigma$ (Fig. \ref{fig:Becerra}(b)), i.e. more weighted towards smaller amplitudes. Bob receives one of 16-QAM signals with $\delta'=\sqrt{\eta}\delta$ and $\sigma'=\sqrt{\eta}\sigma$ due to channel loss. We now find that the Becerra receiver with this modified 16-QAM beats the Gaussian limit under a loss channel with moderate loss ($\eta=0.7$). 
For a fixed $\sigma$, the mutual information first increases and then decreases with $\delta$ ($\bar{n}$). When $\delta$ is too large, only 4 points near the origin contribute to inputs so the mutual information decreases to only $ \log_2 4 = 2 $ bits. The optimal $\sigma$ leading to the maximum mutual information increases with $\bar n$ (inset in Fig. \ref{fig:Becerra}(d)). 
For a small $\bar n$, it is hard to beat the Gaussian limit (squeezed-state scheme). On the other hand, with a moderate value of $\bar n$, the Becerra receiver outperforms both Gaussian schemes (shaded region) attaining a high communication rate ($\gtrsim 3$ bits per channel use) \cite{note:QAMhetero}.

{\it Conclusion}---We identified the capacity limit of general Gaussian settings with multimode Gaussian states and collective Gaussian measurements. We prove the additivity of Gaussian communication, i.e. single-mode communication is optimal without entangled states and joint-measurements. Our finding clarifies the optimal strategy in Gaussian protocols: squeezed-state (coherent-state) scheme in small (large) input-energy regime.
Furthermore, such a fundamental Gaussian benchmark can be used to critically assess the advantages of non-Gaussian receivers in view of capacity. We investigated the Becerra receiver with QAM coherent-states manifesting high communication rate beyond the Gaussian limit remarkably with an appropriate encoding strategy. This seems feasible within current technology considering its recent realization for error rate below SQL.

Our Gaussian benchmark can be very useful in identifying and assessing other non-Gaussian schemes as well. 
One direction to pursue is a joint-measurement receiver showing superadditive capacity like the case of binary inputs \cite{bib:PhysRevA.58.146,bib:PhysRevLett.106.240502}.
We addressed the importance of appropriately choosing an input-state distribution for a given receiver setting to achieve enhanced capacity, which shall be crucially incorporated in future works. 

{\it Acknowledgements}---We acknowledge the support by an NPRP grant 8-352-1-074 from Qatar National Research Fund.

\clearpage

\section{Supplemental Material}

\subsection{Proof for the optimality of separable encoding and decoding}


First we briefly introduce the basics of majorization theory \cite{bib:Majorization} used in our proof. When two $d$-dimensional vectors $\vec{x}$ and $\vec{y}$ sorted in increasing order satisfy the relation $ \sum_{i=1}^{j} x_i \ge \sum_{i=1}^{j} y_i $ for all $ j = 1,2,\cdots,d-1 $ and $ \sum_{i=1}^{d} x_i = \sum_{i=1}^{d} y_i $, we say that $\vec{y}$ majorizes $\vec{x}$, with notation $ \vec{x} \prec \vec{y} $. On the other hand, when the inequality $ \sum_{i=1}^{j} x_i \ge \sum_{i=1}^{j} y_i $ is satisfied for all $ j = 1,2,\cdots,d $, instead of the equality for $j=d$, we say that $\vec{y}$ weakly majorizes $\vec{x}$, with notation $ \vec{x} \prec^{w} \vec{y} $. 

We call a function $f(\vec{x})$ Schur-convex if $ f(\vec{x}) \le f(\vec{y}) $ for all majorized pairs $ \vec{x} \prec \vec{y} $. 
On the other hand, for a weak version of majorization, $ \vec{x} \prec^{w} \vec{y} $ implies $ f(\vec{x}) \le f(\vec{y}) $ if and only if $f$ is Schur-convex and decreasing with respect to all arguments of the function. A symmetric function $f(\vec{x})$, i.e. invariant under permutation of $x_i$ and $x_j$, is provably Schur-convex if and only if $ (x_j-x_i) \left( \frac{\partial f}{\partial x_j} - \frac{\partial f}{\partial x_i} \right) \ge 0 $.
By examining the derivative of the mutual information in Eq. (8) of main text, $ \frac{\partial f}{\partial \lambda_j} = \frac{1}{2}\left( \frac{1}{\nu_\lambda} - \frac{1}{\lambda_j} \right) $, it is straightforward to show that $ f\left( \lambda_1, \lambda_2, \cdots, \lambda_{k_\lambda} \right) $ is decreasing and Schur-convex with respect to the arguments $ 0 \le \lambda_1 \le \lambda_2 \le \cdots \le \lambda_{k_\lambda} \le \nu_\lambda $. 

In the main text, we have introduced CMs $\boldsymbol{\gamma}_\textrm{out} = S_{U_0} \boldsymbol{\gamma}_\textrm{out}^{(D)} S_{U_0}^T$ and $\boldsymbol{\gamma}_\textrm{M} = S_{U_M} \boldsymbol{\gamma}_{M}^{(D)} S_{U_M}^T$ representing the output state and the Gaussian measurement, respectively, at Bob's station. 
For the case of $ S_{U_0} = S_{U_M} = I $, the eigenvalues of $ \boldsymbol{\gamma}_\textrm{out}^{(D)}+\boldsymbol{\gamma}_M^{(D)} $ are simply obtained by the sum of their diagonal elements as $ \vec{\mu} \equiv \frac{1}{2}( e^{-2r_1}+e^{-2s_1}, e^{-2r_2}+e^{-2s_2}, \cdots, e^{-2r_N}+e^{-2s_N}, e^{2r_N}+e^{2s_N}, e^{2r_{N-1}}+e^{2s_{N-1}}, \cdots, e^{2r_1}+e^{2s_1} )^T $ in increasing order.
Denoting the eigenvalues of $\boldsymbol{\gamma}_\textrm{out}+\boldsymbol{\gamma}_\textrm{M}$ by a vector $ \vec{\lambda} \equiv \left( \lambda_1, \lambda_2, \cdots, \lambda_{2N} \right)^T $, we have the majorization relation $ \vec{\lambda} \prec \vec{\mu} $ for any choices of $S_{U_0}$ and $S_{U_M}$. This is because for two Hermitian matrices $X$ and $Y$ with their eigenvalues $\vec{x}$ and $\vec{y}$, respectively, $\vec{x}+\vec{y}$ majorizes the eigenvalue vector of $X+Y$ \cite{bib:MatrixAnalysis}.
Even though we have above shown that the mutual information in Eq. (8) of main text is Schur-convex, the relation $ \vec{\lambda} \prec \vec{\mu} $ does not immediately guarantee 
	\begin{equation} \label{eq:mimajorization}
	f\left( \lambda_1, \lambda_2, \cdots, \lambda_{k_\lambda} \right) \le f\left( \mu_1, \mu_2 \cdots, \mu_{k_\mu} \right),
	\end{equation} 
as the left- and the right-hand sides of Eq. (\ref{eq:mimajorization}) may have different number of arguments, i.e. $k_\lambda\neq k_\mu$. Nevertheless, we now show that the inequality (\ref{eq:mimajorization}) is valid. Recall that $k_\mu$ and $k_\lambda$ are the numbers of nonzero signal assigned by water-filling solution to the noise vectors $\vec{\mu}$ and $ \vec{\lambda}$, respectively. 

(i) For the case of $ k_\mu = k_\lambda $,  $ \vec{\lambda} \prec \vec{\mu} $ directly implies the weak majorization relation of truncated vectors, $ \left( \lambda_1, \lambda_2, \cdots, \lambda_{k_\lambda} \right)^T \prec^w \left( \mu_1, \mu_2, \cdots, \mu_{k_\mu} \right)^T $ and we thus find $ f\left( \lambda_1, \lambda_2, \cdots, \lambda_{k_\lambda} \right) \le f\left( \mu_1, \mu_2 \cdots, \mu_{k_\mu} \right) $. 
   \begin{figure}[t!]
\centering \includegraphics[width=0.85\columnwidth]{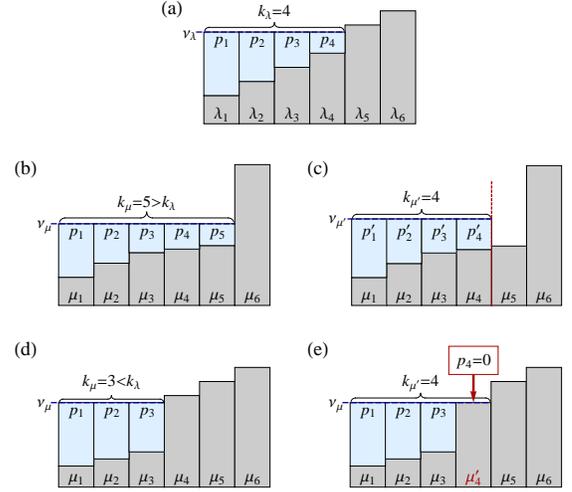}
	\caption{\label{fig:water} Illustration of water-filling solutions (optimal encoding) for two different noise vectors (a) $\vec{\lambda}$ (case of generalized inputs and decoding) and (b,d) $\vec{\mu}$ (case of separable inputs and decoding). For the case of $ k_\mu > k_\lambda $, we show (b) the optimal water-filling solution for a noise vector $\vec{\mu}$ and (c) an alternative encoding scheme assigning only $ k_{\mu'} = k_\lambda $ number of signals. For the case of $ k_\mu < k_\lambda $, we show (d) the optimal water-filling solution and (e) an encoding achieving the same capacity while the noise is adjusted so that $ k_{\mu'} = k_\lambda $ including null signal $ s_j = 0 $.}
	\end{figure}
	
(ii) When $ k_\mu > k_\lambda $, let us consider an alternative encoding where we only assign $k_\lambda$ signal to the noise vector $ \vec{\mu} $ [Fig. \ref{fig:water}(c)], instead of the optimal encoding with $k_\mu$ signal [Fig. \ref{fig:water}(b)]. Under this restriction, the mutual information becomes maximal when the signal is assigned by water-filling method to the noise vector $ \left( \mu_1, \mu_2, \cdots, \mu_{k_\lambda} \right)^T $ as depicted in Fig. \ref{fig:water}(c). Note that the new ``water level" $\nu_{\mu'}$ is higher than $\nu_\mu$ due to the energy constraint. Of course, this is not the optimal strategy for the given noise vector $\vec{\mu}$, however, we have $ f\left( \lambda_1, \lambda_2, \cdots, \lambda_{k_\lambda} \right) \le f\left( \mu_1, \mu_2 \cdots, \mu_{k_\lambda} \right) $ due to the majorization relation $ \left( \lambda_1, \lambda_2, \cdots, \lambda_{k_\lambda} \right)^T \prec^w \left( \mu_1, \mu_2, \cdots, \mu_{k_\lambda} \right)^T $. It thus gives the desired relation $ f\left( \lambda_1, \lambda_2, \cdots, \lambda_{k_\lambda} \right) \le f\left( \mu_1, \mu_2 \cdots, \mu_{k_\lambda} \right) \le f\left( \mu_1, \mu_2 \cdots, \mu_{k_\mu} \right) $.

(iii) When $ k_\mu < k_\lambda $, we define a $k_\lambda$-dimensional vector $ \vec{\mu}' = \left( \mu_1, \mu_2, \cdots, \mu_{k_\mu}, \nu_\mu, \nu_\mu, \cdots, \nu_\mu \right)^T $, where $\nu_\mu$ is the water level for the original vector  $\vec{\mu}$. Because $ \mu_j \ge \nu_\mu$ for $ j > k_\mu $, we have the majorization relation $ \left( \lambda_1, \lambda_2, \cdots, \lambda_{k_\lambda} \right)^T \prec^w \left( \mu_1, \mu_2, \cdots, \mu_{k_\lambda} \right)^T \prec^w \vec{\mu}' $. Using the noise vector $ \vec{\mu}' $, the optimal encoding is the same as that for $\vec{\mu}$ [Fig. \ref{fig:water} (e)], thus the mutual information does not change. In fact, we only add {\it null} signal $p_j=0$ for $ j = k_\mu+1, k_\mu+2, \cdots, k_\lambda $ [Fig. \ref{fig:water}(e)]. Because we set elements of $\vec{\mu}'$ such that $ 0 \le \mu_1' \le \mu_2' \le \cdots \le \mu_{k_\lambda}' \le \nu_\mu $, $f(\vec{\mu}')$ is again decreasing and Schur-convex within this range. It thus gives the desired relation $ f\left( \lambda_1, \lambda_2, \cdots, \lambda_{k_\lambda} \right) \le  f\left( \vec{\mu}' \right) = f\left( \mu_1, \mu_2 \cdots, \mu_{k_\mu} \right) $.

\subsection{Gaussian communication schemes}
A coherent-state scheme employs symmetric two-quadrature encoding on a coherent-state input and balanced heterodyne detection, that is, $ \boldsymbol{\gamma}_\textrm{in} = \frac{1}{2}I $, $ P_\textrm{in} = \bar{n}I $, and $ \boldsymbol{\gamma}_M = \frac{1}{2}I $. Its capacity is given by
	\begin{equation}
	C^\textrm{coh} = \log_2 \left( 1+ \frac{2\tau\bar{n}}{1+\tau+2m} \right) .
	\end{equation}
Squeezed-state scheme employs single-quadrature encoding on a squeezed-state input and homodyne detection, that is, $ \boldsymbol{\gamma}_\textrm{in} = \frac{1}{2}\textrm{diag}\left( e^{-2r}, e^{2r} \right) $, $ P_\textrm{in} = \textrm{diag}\left( 2(\bar{n}-\sinh^2 r), 0 \right) $, and $ \boldsymbol{\gamma}_M = \frac{1}{2}\textrm{diag}\left( e^{-2s}, e^{2s} \right) $ with $ s \to \infty $. With the choice of optimal squeezing $ \exp(2r) = \frac{-\tau+\sqrt{8\tau m\bar{n} + (\tau+2m)^2}}{2m} $ \cite{bib:PhysRevA.50.3295,bib:OptCommun.149.152,bib:PhysRevA.91.042336}, the capacity of squeezed-state scheme becomes
	\begin{equation}
	C^\textrm{sq} = \log_2 \left( \frac{-\tau+\sqrt{8\tau m\bar{n} + (\tau+2m)^2}}{2m} \right) .
	\end{equation}

\begin{figure}
	\centering \includegraphics[width=\columnwidth]{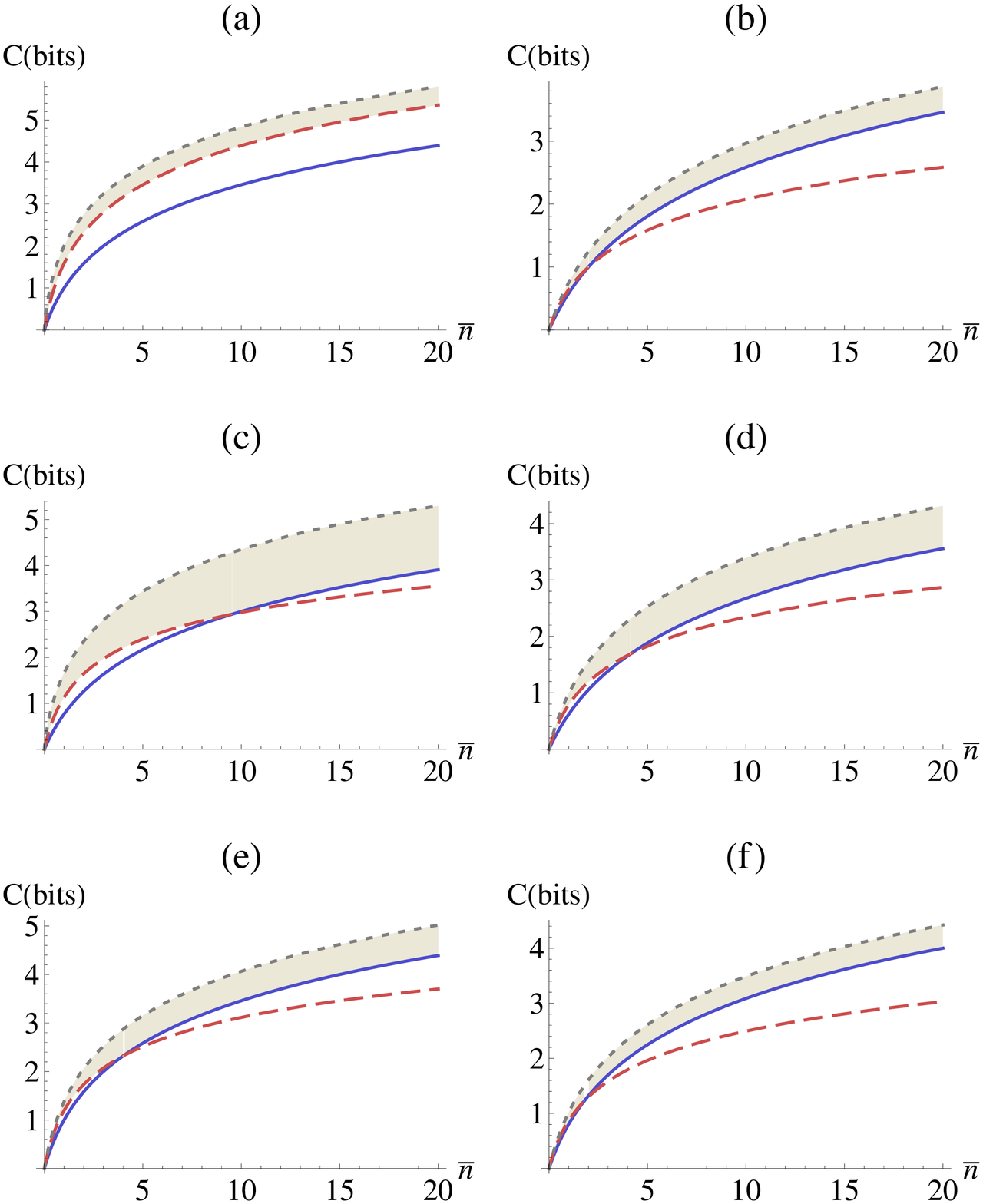}
	\caption{\label{fig:capacity} Plot comparing the capacity of coherent-state scheme (blue solid curves), that of squeezed-state scheme (red dashed curves), and the Holevo bound (gray dotted curves) for different types of channel (a) an ideal channel, (b) a channel with added noise $m=1$, (c) a pure-loss channel with $\tau=0.5$ and $n_\textrm{th}=0$, (d) a loss channel with $\tau=0.7$ and $n_\textrm{th}=1$, (e) a quantum-limited amplification channel with $\tau=1.5$ and $n_\textrm{th}=0$, and (f) an amplification channel with $\tau=1.5$ and $n_\textrm{th}=1$. The shaded region represents the gap between the Holevo bound and the capacity of optimal Gaussian communication.}
	\end{figure}

In Fig. \ref{fig:capacity}, we show the capacities of two Gaussian schemes together with the Holevo bound \cite{bib:NatPhoton.8.796,bib:NatPhoton.7.834} given by
	\begin{eqnarray}
	C^\textrm{Holevo} & = & g \left( \tau\bar{n} + m+{\textstyle \frac{\tau-1}{2}} \right) - g \left( m+{\textstyle \frac{\tau-1}{2}} \right) , \\
	g(x) & \equiv & (1+x)\log_2(1+x) - x\log_2 x , \nonumber
	\end{eqnarray}
where $g(n)$ is the von Neumann entropy of a thermal state with thermal photon number $n$.
We find that there exists a gap (shaded region) between the Holevo bound and the capacity of optimal Gaussian communication. For an ideal channel [Fig. \ref{fig:capacity}(a)], the squeezed-state scheme always beats the coherent-state scheme \cite{bib:RevModPhys.66.481} and attains the capacity close to the Holevo bound. However, as some noise is added [Fig. \ref{fig:capacity}(b)], the capacity of the squeezed-state scheme grows with energy $\bar n$ less prominently than the coherent-state scheme, and the coherent-state scheme instead attains the capacity close to the Holevo bound. Except for the ideal case, there always exists a crossover between $C^\textrm{coh}$ and $C^\textrm{sq}$. This is because a squeezed state is more fragile against the channel noise than a coherent state \cite{bib:NatPhoton.8.796,bib:NatCommun.5.3826}. One can readily find that the coherent-state scheme is optimal when $ \bar{n} \ge \frac{1+2m+\tau}{2m\tau} $, and otherwise, the squeezed-state scheme is optimal. For the loss channel and the amplification channel with $n_\textrm{th}=0$, we find a small-energy region where the squeezed-state scheme beats the coherent-state scheme [Fig. \ref{fig:capacity}(c,e)], but this region becomes smaller as $n_\textrm{th}$ increases [Fig. \ref{fig:capacity}(d,f)]. The gap between the Holevo bound and the capacity of coherent-state scheme decreases as $n_\textrm{th}$ increases, or $m$ increases.

\subsection{QAM encoding with heterodyne detection}

\begin{figure}[!t]
	\centering \includegraphics[width=\columnwidth]{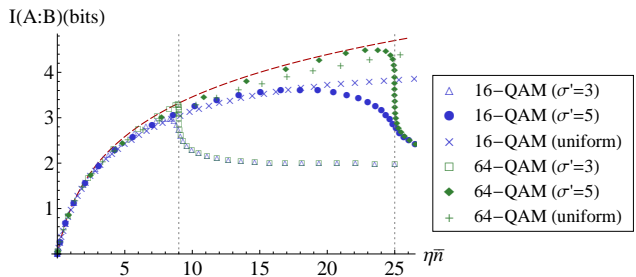}
	\caption{\label{fig:QAM} Mutual information attained with 16(64)-QAM and balanced heterodyne detection. Each point is obtained numerically with different encoding strategies as represented in legends and with different lattice spacing $\delta$. The dashed red curve shows the capacity of coherent-state scheme $C^\textrm{coh}$ and vertical dotted lines represent $ \eta\bar{n} = \sigma'^2 $ for $ \sigma' = \{3,5\} $, respectively.}
	\end{figure}

We here show that it is possible to approach the capacity of coherent-state scheme obtained under the condition of continuous modulation by using a {\it finite} number of coherent states for encoding.
In Fig. \ref{fig:QAM}, we plot the mutual information attained with 16(64)-QAM and balanced heterodyne detection under a pure-loss channel with transmittance $\eta$, along with the capacity of coherent-state scheme $ C^\textrm{coh} = \log_2 (1+\eta\bar{n}) $ based on continuous modulation. 
We see that the mutual information attained using the $M$-QAM with Gaussian-like distribution (more weighted towards the origin of phase space) approaches very closely the capacity of coherent-state scheme (optimal strategy with heterodyne receiver) under the same energy, with the condition $ \eta\bar{n} \lesssim \sigma'^2 $. The maximum mutual information under $M$-QAM is given by $\log_2 M$ bits, i.e., 4(6) bits for 16(64)-QAM, and is achieved when the points are perfectly distinguishable. 
The mutual information drops as the spacing $\delta$ becomes large enough as $ \eta\bar{n} > \sigma'^2 $ (vertical dashed lines). When $\delta$ is too large, only 4 points near the origin become significant among the inputs so  the mutual information decreases to only $ \log_2 4 = 2 $ bits. On the other hand, if we employ $M$-QAM with uniform distribution ($ \sigma \to \infty $), $I(A:B)$ shows a prominent gap from $C^\textrm{coh}$ with increasing energy.

\begin{thebibliography}{1}
\bibitem{note:optimalsingle} Although the proof in Ref. \cite{bib:PhysRevA.91.042336} was given only for loss channels with $ 0 < \tau < 1 $, it is straightforward to extend to all phase-insensitive Gaussian channels.
\bibitem{note:QAMtransmission} Under a pure-loss channel, the capacity of a scheme employing coherent-state inputs is given as a function of output photon number $\eta\bar{n}$. This is because a pure-loss channel transforms a coherent state into another coherent state with only amplitude decreased by $\eta$.
\bibitem{note:QAMhetero} As another practical issue, it is not possible to implement a truly continuous modulation of input states for encoding, and one prepares only a finite (however large) number of coherent states. It is then interesting to know whether a scheme based on finite inputs can attain capacity emerging under the condition of continuous modulation, e.g. Holevo capacity and Gaussian communication capacity. In \cite{supple}, we show that our modified QAM, i.e. using a finite number of coherent states, can achieve the capacity of Gaussian communication by using a Gaussian receiver (heterodyne detection).
\end{thebibliography}
\end{document}